\documentstyle[12pt]{article}

\def\be{\begin{equation}}
\def\ee{\end{equation}}
\def\a{\alpha}

\def\g{\gamma}

\def\ln{\mbox{ln}}

\def\cos{\mbox{cos}}
\def\sin{\mbox{sin}}
\def\sh{\mbox{sh}}

\def\ra{\rangle}

\def\Im{\mbox{Im}}
\def\tp{t^{\prime}}
\def\sign{\mbox{sign}}

\begin{document}

\begin{center}
{\bf On the particle excitations in the XXZ- spin chain.} 
\end{center}
\vspace{0.2in}
\begin{center}
{\large A.A.Ovchinnikov}  
\end{center}

\begin{center}
{\it Institute for Nuclear Research, RAS, Moscow}
\end{center}

\vspace{0.1in}

\begin{abstract}

  We continue to study the excited states for the XXZ- spin chain 
corresponding to the complex roots of the Bethe Ansatz equations 
with the imaginary part equal to $\pi/2$. 
  We propose the particle-hole symmetry which relates the 
eigenstates build up from the two different pseudovacuum states. 
  We find the XXX- spin chain limit for the eigenstates with the 
complex roots. 
  We also comment on the low-energy excited states for the 
XXZ- spin chain.

\end{abstract}

\vspace{0.2in}

{\bf 1. Introduction}

\vspace{0.2in}

In the present paper we study the excited states of the XXZ- spin chain. 
The Bethe Ansatz equations \cite{Bethe} for the XXZ- spin chain have the 
following form 
\be
\left(\frac{\sh(t_{\a}-i\eta/2)}{\sh(t_{\a}+i\eta/2)}\right)^{L} = 
\prod_{\g\neq\a}\frac{\sh(t_{\a}-t_{\g}-i\eta)}{\sh(t_{\a}-t_{\g}+i\eta)}, 
\label{BA}
\ee
where $t_{\a}$, $\a=1,\ldots M$ are the complex parameters (roots) which 
determine the Bethe wave function and the anisotropy parameter 
$\Delta=\cos(\eta)$. It is convenient to define the following functions: 
\[
\phi(t)=\frac{1}{i}\ln\left(-\frac{\sh(t-i\eta/2)}{\sh(t+i\eta/2)}\right),~~
\phi_{2}(t)=\frac{1}{i}\ln\left(-\frac{\sh(t-i\eta)}{\sh(t+i\eta)}\right). 
\] 
Then the Bethe Ansatz equations take the following form:
\be
L\phi(t_{\a})=2\pi n_{\a}+\sum_{\g\neq\a}\phi_2(t_{\a}-t_{\g}), 
\label{n}
\ee
where the quantum numbers $n_{\a}$ are integers for $M$- odd and half-integers 
for $M$- even (we assume the length of the chain $L$ to be even).  
For the ground state at the fixed magnetization $S^{z}=M-L/2<0$ 
all the parameters $t_{\a}$ are real \cite{O}, while for the excited 
states the part of the parameters $t_{\a}$ can be the complex numbers. 
It is well known that for the excited states the complex roots 
appear in the form of the so-called $k$- strings \cite{Takahashi}. 
These complex roots are studied in greate detail in the literature. 
    However, it is obscure in the literature that the complex roots can 
appear also in the form $t_{\a}=t_{\a}^{\prime}+i\pi/2$, where 
$t_{\a}^{\prime}$- is real. These roots constitute the upper half of the 
Brillouin zone for the particles in the XXZ- spin chain.  
In contrast to the holes we refer to these roots as particles 
throughout the paper and did not consider the other complex roots.  
Not long ago the excited states of this type have been found for the 
more general case of the Bethe Ansatz equations for the Sine-Gordon 
model \cite{V}.  
However in this paper some questions did not received the 
considerable attention. 
  In particular, the limiting case of the XXX- spin chain was not 
considered. 
   In the present paper 
  we continue to study the excited states for the XXZ- spin chain 
corresponding to the complex roots of the Bethe Ansatz equations 
with the imaginary part equal to $\pi/2$. 
  We propose the particle-hole symmetry which relates the 
eigenstates build up from the two different pseudovacuum states. 
  We find the XXX- spin chain limit for the eigenstates with the 
complex roots. 
  We also comment on the low-energy excited states for the 
XXZ- spin chain.

Let us explain why an arbitrary number of complex roots of the form 
$t_{\a}=t_{\a}^{\prime}+i\pi/2$, where $t_{\a}^{\prime}$- is real,  can exist. 
First, consider the limit of the XX- spin chain ($\eta=\pi/2$) where the 
system is equivalent to the free fermions. 
From the equations $L\phi(t_{\a})=n_{\a}$ we see that for $|n_{\a}|<L/4$ 
the root $t_{\a}$ is real while for $|n_{\a}|>L/4$ the root $t_{\a}$ is  
complex: $t_{\a}=t_{\a}^{\prime}+i\pi/2$, where $t_{\a}^{\prime}$- is real. 
There is the statement that any complex root of the equations (\ref{BA}) 
is accompanied by its complex conjugate. This theorem is not applicable 
to the complex roots considered in the present paper since the ratios in 
the equations (\ref{BA}) do not change when the complex conjugate root 
is substituted. However clearly this theorem shows that these complex roots 
remain at the line $\Im t_{\a}=\pi/2$ when $\eta$ is decreased. 
Second, one can define the function $n_L(t)$ (see eq.(\ref{nl}) below) 
obeying the condition $n_L(t_{\a})=n_{\a}$. Thus formally the root $t_{\a}$ 
is the solution of the equation $n_L(t)=n_{\a}$. Analysing the function 
$n_L(t)$ we find that $t_{\a}$ is real when $|n_{\a}|<n_c$, while 
$t_{\a}=t_{\a}^{\prime}+i\pi/2$ ($t_{\a}^{\prime}$ is real) when $|n_{\a}|>n_c$, 
where $n_c$ is some number which depends on $\eta$ and $M$. 
In particular at $M=L/2$ we have the boundary $n_c=L/4$. 
Note that equally well one can consider the dependence of 
some quantum number, say $n_1$, on the parameter $t_1$ in the equations 
(\ref{n}) at the fixed roots $t_{\a}$, $\a\neq 1$. Evaluating the functions 
$\phi(t)$, $\phi_2(t)$ for the complex arguments we obtain that at 
$M=L/2$ at $|n_1|>L/4$ we have the complex solutions $t_1=t_1^{\prime}+i\pi/2$ 
which leads us to the same conclusion.
Thus the existence of the complex roots at the axis $\Im t=\pi/2$ is 
demonstrated.

\vspace{0.2in}

{\bf 2. Particle-hole symmetry}

\vspace{0.2in}

Here we introduce the new type of symmetry for the XXZ- spin chain. 
Namely we argue that there is an exact equality between the eigenstates 
obtained starting from the two different pseudovacuum states 
(spin-up and spin-down states) when the corresponding quantum numbers 
are related in a simple way. The equations of this type is the 
manifestation of the so called particle-hole symmetry of the XXZ- spin 
chain, 

It is well known (for example see \cite{FT}) that for the isotropic 
XXX- spin chain the Bethe eigenstates with $M>L/2$ are exactly equal to 
zero although the solutions of the Bethe Ansatz equations can exist,  
which is connected with the conservation of the total spin of the system.  
This is not the case for the general XXZ- spin chain. 
In this case the solutions of the equations (\ref{BA}) at $M>L/2$ can 
exist and the corresponding eigenstate is not zero. 
Now we observe that each eigenstate for $M<L/2$ can be obtained in the 
two different ways: either starting from the usual pseudovacuum state 
with all spins directed down or starting from the different pseudovacuum 
states with all spins directed up. In the last case we have the number 
of the roots $t_{\a}^{\star}$ obeying the same BA equations (\ref{BA}) 
$M^{\star}=L-M>L/2$. The eigenstate is determined by the set of the 
quantum numbers $n_{\a}$, $\a=1\ldots M$, while the same eigenstate 
constructed from the different pseudovacuum corresponds to the different 
quantum numbers $n_{\a}^{\star}$, $\a=1\ldots M^{\star}$. 
The so called particle-hole symmetry gives us the relation between the 
two sets $\{n_{\a}\}$ and $\{n_{\a}^{\star}\}$. 
In fact let us denote by $n_{0i}$ the set of the quantum numbers 
corresponding to the holes in the state given by the set $\{n_{\a}\}$ 
(all numbers that are absent in the set $\{n_{\a}\}$ on the interval 
$(-L/2,L/2)$). Then the set $\{n_{\a}^{\star}\}$ corresponding to the 
``dual'' state is given by the following formulas:  
\be
n_i^{\star}=L/2-n_{0i}, ~~~(n_{0i}>0),~~~~~~
n_i^{\star}=-L/2-n_{0i}, ~~~(n_{0i}<0),
\label{S}
\ee
where $i=1,\ldots M^{\star}$, $M^{\star}=L-M$.  
The equations (\ref{S}) are the exact relations between the quantum 
numbers corresponding to the eigenstates obtained starting from the 
two different pseudovacuum states. 
The relations (\ref{S}) for the XXZ- spin chain follow from 
the relations for the XX- spin chain ($\Delta=0$) which can be solved 
using the mapping of the spin operators to the fermionic operators 
(Jordan-Wigner transformation). Using the continuity argument one 
see that the relations (\ref{S}) are valid for an arbitrary $\Delta$ 
(XXZ- spin chain) as well. 
In fact varying $\Delta$ at fixed quantum numbers we see that 
two eigenstates coincide at arbitrary $\Delta$. 
We will see below what is going on with the relations (\ref{S})  
for the case of the XXX- spin chain ($\Delta=1$) where the particle 
(complex roots) excitations does not exist and for the Bethe eigenstates 
we always have $M<L/2$.

\vspace{0.2in}

{\bf 3. Evolution of the eigenstates with particles}

\vspace{0.2in}

Let us consider the evolution of the eigenstates with particles 
(complex roots) when we vary the parameter $\eta$ from 
$\eta=\pi/2$ (XX- spin chain) to $\eta=0$ (XXX- spin chain). 
The evolution of this kind was first studied numerically in \cite{V} 
in the context of the Bethe Ansatz equations for the Sine-Gordon model. 
Below we present the simplified picture which is based on the 
assumption that the so called counting function (see below) 
is the monotoneusly increasing function of its argument. 
Although actually this is not always the case at the left and the right 
ends of the distribution of real roots for our purposes 
it is sufficient to present here the simplified picture of the 
evolution reffering to ref.\cite{V} for more details.  
First we present this simplified picture and then we briefly 
explain how the effects of non- monotonousless \cite{V} modify the 
results. 
For example, consider the eigenstate without the holes on the 
real axis and some number of complex roots, such that the total 
number of roots 
\be
M=\frac{L}{2}+k 
\label{M}
\ee
fulfills the condition $M>L/2$ ($k>0$). Let us denote symbolically 
by $|\phi(R,C)\ra$ the eigenstate with $R$ real roots (without holes) 
and $C$ complex roots on the axis $\Im(t)=\pi/2$ ($t=\tp+i\pi/2$, 
$\tp$- is real) corresponding to some quantum numbers 
(in the following we did not write down these quantum numbers explicitly).  
Let us determine the allowed values of $R$ and $C$ for a given 
value of the parameter $\eta$. 
In order to find the number of real roots $R$ it is convenient to 
introduce the following function (see \cite{V} and references therin) 
\be
n_L(t)=\frac{1}{2\pi}\left(L\phi(t)-\sum_{\g}\phi_2(t-t_{\g})\right),  
\label{nl}
\ee
where the sum at the right-hand side is over all roots. 
For our purposes it is sufficient to assume that the function (\ref{nl}) 
is the monotoneously increasing function of its argument. 
Due to the equations (\ref{n}) this function obeys the following remarkable 
property: $n_L(t_{\a})=n_{\a}$. In order to find $R$ one can find the 
maximal allowed quantum number $n_{max}$. To find this number it is 
sufficient to calculate the critical value $n_{c}=n_L(\infty)$. 
We use the following limiting values of the functions $\phi(t)$, $\phi_2(t)$: 
$\phi(\pm\infty)=\pm(\pi-\eta)$, $\phi_2(\pm\infty)=\pm(\pi-2\eta)$. 
For example, consider the case $k$- odd. In this case the quantum numbers 
$n_{\a}$ are integers and we obtain $n_{max}=[n_c]$, where the parenthesis 
mean the integer part. Then for the allowed value of the number of real roots 
$R=2n_{max}+1$ we finally obtain the equation: 
\be
R=\frac{L}{2}-k+2\left[\frac{1}{2}+\frac{k\eta}{\pi}\right]. 
\label{R}
\ee
It is easy to prove that for $k$- even we obtain exactly the same expression. 
The number of the complex roots 
\be
C=M-R=2k-2\left[\frac{1}{2}+\frac{k\eta}{\pi}\right]. 
\label{C}
\ee
Thus we obtain the following eigenstate without the holes on the real axis: 
\be
|\phi(R,C)\ra=|\phi\left(\frac{L}{2}-k+2\left[\frac{1}{2}+\frac{k\eta}{\pi}\right],~~
2k-2\left[\frac{1}{2}+\frac{k\eta}{\pi}\right]\right)\ra. 
\label{state}
\ee
In the same way at $M<L/2$ or $k=-|k|<0$ (see eq.(\ref{M})) we obtain the similar 
expression for the number of vacancies on the real axis: 
\be
R=\frac{L}{2}+|k|-2\left[\frac{1}{2}+\frac{|k|\eta}{\pi}\right].
\label{Rk}
\ee
In particular one can see from eq.(\ref{Rk}) that the number of holes 
is always even in the general case of the XXZ- spin chain 
(in the case of the XXX- spin chain this fact was observed in ref.\cite{FT}). 
Now let us proceed with studying the eigenstate (\ref{state}).  
Let us vary the parameter $\eta$ from $\eta=\pi/2$ to $\eta=0$ at fixed 
quantum numbers. For the XX- spin chain, $\eta=\pi/2$, 
the eigenstate (\ref{state}) takes the form $|\phi(L/2,k)\ra$, so that we have 
$L/2$ particles with the momenta $|k_{\a}|<\pi/2$ and $k$ particles with the 
momenta $|k_{\a}|>\pi/2$, while for $\eta\rightarrow 0$ 
the eigenstate (\ref{state}) takes the form $|\phi(L/2-k,2k)\ra$. 
That means that the roots disappear from the real axis and appear on the 
axis $\Im t=\pi/2$. From eq.(\ref{state}) one can see that the first jump 
takes place at $\eta=\pi/2k$ where the number of complex roots increases by $2$,
and the next jumps occur at $\eta_{m}=(\pi/k)(m+1/2)$, $m\geq 0$. 
Clearly, first the largest root at the real axis goes to the infinity, 
and then jumps to the infinity at the axis $\Im t=\pi/2$. 
     We have the following general picture. The jumps of the real roots to the 
line $\Im t=\pi/2$ occur because at the given $\eta$ and $k$ the 
real roots $t_{\a}$ correspond to the quantum numbers $|n_{\a}|<n_c$, 
while the complex roots $t_{\a}=t_{\a}^{\prime}+i\pi/2$ 
correspond to the quantum numbers $|n_{\a}|>n_c$. 
Since the boundary $n_c$ depends on $\eta$ and $k$ at $k\neq 0$ the variation 
of $\eta$ at the fixed quantum numbers leads to the appearance of the 
  new complex roots. 
     Let us note that at infinity at the axis $\Im t=\pi/2$ the function 
$n_L(t)$ takes exactly the same value $n_c$ as at the real axis. 
Clearly, the jumps of the real roots occur when the boundary 
$n_c= n_L(\infty)$ becomes an integer or half-integer 
and the corresponding maximal root is pushed to the infinity. 
One can equally well consider the initial eigenstate of the 
XX- spin chain which contains both the particles (complex roots) 
and the holes in the distribution of roots on the real axis.  
Clearly when $\eta$ decreases the holes will also jump to 
the line $\Im t=\pi/2$ at the same points. 
Then the number $R$ will be less than that in the state (\ref{state}) 
by the number of holes that have not jumped to the line $\Im t=\pi/2$ 
in the process of the evolution. 

For example consider now the case $k=-|k|<0$. Since now the value 
of $n_c$ increases with decreasing of $\eta$ (see eq.(\ref{Rk})) 
the roots and the holes jump down from the line $\Im t=\pi/2$ to 
the real axis. Clearly, if there are no complex roots in the 
initial state, then they cannot appear in the final state. 
If in the XX- spin chain limit there are both the particles 
and the holes, then in the final state we still have $L/2+|k|$ 
real vacancies and $M=L/2-|k|$ roots. Then at the real axis 
we have the holes of the two types and the part of the particles 
that have jumped downwards. Clearly, the remaining particles 
(complex roots) are transform into $S^{+}$- operators in the 
limit of the XXX- spin chain (see the next Section).

The picture described above is exactly valid for the jumps of the holes 
since in this case $n_L(t)$ is in fact globally monotonic. 
Numerical calculations \cite{V} shows that in the case of particles 
the function $n_L(t)$ is not monotonic in the left and the right tails 
of the distribution which is important sufficiently close to the points 
$\eta_{m}$. In the vicinity of these points there is rather complicated 
behaviour of the roots and the special holes (see \cite{V}).  
This leads to the shifts of the points of jumps 
from the points $\eta_{m}$ to some configuration- dependent points 
close to the points $\eta_{m}$ \cite{V}. 
Note that in the vicinity of these points the equation (\ref{R}) 
for the number of vacancies is violated.

\vspace{0.4in}

{\bf 4. XXX- limit of the states with particles}

\vspace{0.2in}

The XXX- spin chain does not have the complex roots described 
in the present paper. Let us determine the eigenstates of the 
XXX- spin chain which are the final result of the evolution 
at $\eta\rightarrow 0$ of the states (\ref{state}) of the 
XXZ- spin chain. At arbitrary $k$ and $\eta\rightarrow 0$ the 
eigenstate (\ref{state}) takes the form $|\phi(L/2-k,2k)\ra$, 
where $L/2-k$ real roots occupy all vacancies on the real axis, 
and $2k$ complex roots correspond to some quantum numbers $n_i$, 
$|n_i|>L/4-k/2$, which are not indicated explicitly. 
We observe that at $\eta\rightarrow 0$ all the real roots are of 
order of $\eta$ (the maximal root is $\sim\eta\ln L$) , 
while the complex roots are at finite distance from zero. 
One can see that at $\eta\rightarrow 0$ the operator $B(t)$, 
which creates particles in the framework of the Algebraic 
Bethe Ansatz method, behaves as 
\[ 
B(t)|_{t-fixed}\rightarrow \eta f(t)S^{+}+O(\eta^2), 
\]
where $f(t)$ is some function. That means that at $\eta=0$ all 
complex roots lead to the operators $S^{+}$ acting on the 
eigenstates of the XXX- spin chain. The same phenomenon can be 
predicted in the framework of the Coordinate Bethe Ansatz 
taking into account the behaviour of the function 
$\tilde{\phi}(t)=\phi(t+i\pi/2)$ at $\eta\rightarrow 0$. 
In fact in this limit the function $\tilde{\phi}(t)$ takes the 
form $\tilde{\phi}(t)=\pi\sign(t)$ and the momenta 
$k_{\a}=\phi(t_{\a})$ in the Coordinate Bethe Ansatz corresponding 
to the complex roots take the values $\pm\pi$. 
Thus we obtain: 
\be
|\phi(L/2-k,2k)\ra\rightarrow (S^{+})^{2k}|\phi(L/2-k)\ra_{XXX}^{(2k)}.
\label{lim}
\ee
The eigenstate of the XXX- spin chain at the right-hand side of 
eq.(\ref{lim}) contains $L/2-k$ roots. That means that the number of 
vacancies for the real roots for the XXX- spin chain is equal to 
$L/2+k$. That means that there is $2k$ holes in this state, which 
is indicated in eq.(\ref{lim}). Clearly, the positions of these $2k$ 
holes are determined by the positions of $2k$ complex roots in the 
state at the left-hand side of the equation (\ref{lim}). 
Namely, the quantum numbers of the holes are: $n_{0i}=L/2-n_i$ ($n_i>0$), 
$n_{0i}=-L/2-n_i$ ($n_i<0$). 
One can come to the same conclusion using the following arguments. 
Using the particle-hole symmetry (see Section 2) one can see that 
the eigenstate at the left-hand side of the equation (\ref{lim}) in the 
limit $\eta\rightarrow 0$ takes the form 
$|\phi(L/2-k)\ra_{XXX}^{(2k)(\star)}$ with the holes at the positions 
$n_{0i}$. At the same time it is clear that this state coincide with the 
eigenstate at the right-hand state of eq.(\ref{lim}), since 
the states  $|\phi(L/2-k)\ra_{XXX}^{(2k)}$ and 
$|\phi(L/2-k)\ra_{XXX}^{(2k)(\star)}$ belong to the same spin multiplet. 
Thus we come to the conclusion that the equation (\ref{lim}) is true. 
We find that the equation (\ref{lim}) follows from the particle-hole 
symmetry.
     Since in the presence of the complex roots the Bethe Ansatz 
equations (\ref{n}) for the XXZ- spin chain do not become the 
Bethe Ansatz equations for the XXX- spin chain in the limit 
$\eta\rightarrow 0$, the quantum numbers for the the XXZ- spin chain 
and the quantum numbers for the state (\ref{lim}) are different. 
In fact let us denote by $n_i$ the quantum numbers corresponding to the 
complex roots for the XXZ- spin chain close to the point $\eta=0$. 
Let us denote by $t_{i}^{\prime}$ the real parts of the corresponding 
complex roots. Then for the real roots $t_{\a}$ 
the Bethe Ansatz equations for the XXZ- spin chain 
close to the point $\eta=0$ take the form: 
\be
L\phi(t_{\a})=2\pi n_{\a}+\sum_{\g\neq\a}\phi_2(t_{\a}-t_{\g}) 
+\sum_{i\in C}f(t_{\a}-t_{i}^{\prime}),  
\label{2}
\ee
where the functions $\phi(t)$, $\phi_{2}(t)$ correspond to the 
functions close to the point of the XXX- 
spin chain and the function $f(t)=\pi\sign(t)$. 
Here the quantum numbers $n_{\a}$ correspond to the state with no holes.
From the equation (\ref{2}) one can see that there is the holes 
in the eigenstate of the XXZ- spin chain which turn to the eigenstate 
(\ref{lim}) in the limit $\eta\rightarrow 0$ 
at the positions related to the XXZ- spin chain parameters 
$t_{0i}=t_{i}^{\prime}$. Then the positions of the holes 
$n_{0i}$ corresponding to the parameters $t_{0i}$  
in the XXZ- spin chain obey the 
condition $n_{0i}=L/2-n_i$ ($n_{i}>0$) in agreement with the predictions 
    of the particle-hole symmetry (\ref{S}). 
Using the Fourier transforms of the derivatives of the functions 
$n_L(t)$ defined for the real roots ($n_L(t_{0i})=n_{0i}$)
and $\tilde{n}_L(t)=n_L(t+i\pi/2)$ ($\tilde{n}_L(t_i^{\prime})=n_i$) 
one can confirm these relations explicitly up to the terms of order of $O(1)$. 
Thus one can see how the holes in the eigenstate of the XXX- spin chain 
in the equation (\ref{lim}) can appear. 
Thus we see how the eigenstates of the XXZ- spin chain with particles 
(complex roots) transform smoothly into the eigenstates of the XXX- spin 
chain in the limit $\eta\rightarrow 0$.

Let us describe the low-energy 
excited states of the XXZ- spin chain 
corresponding to the small values of $k$ (\ref{M}). 
First consider the case $k=1$ which corresponds to the one 
extra particle (complex root). 
However in this case according to the equation (\ref{R}) the number 
of vacancies on the real axis become $L/2-1$ which means that 
one extra real root becomes complex. As a result we have the state 
with two complex roots (two particles) and $L/2-1$ real roots. 
The dispersion relation corresponding to the one particle 
(complex root) can be obtained in the same way as for the one hole 
(for example see \cite{G}). The calculations show that the 
dispersion relation is identical to the dispersion relation of one
hole:
\be 
\epsilon(p)=\frac{\pi}{2}\frac{\sin(\eta)}{\eta}\sin(p). 
\label{e}
\ee
The total energy and the momentum are 
$\Delta E=\epsilon(p_1)+\epsilon(p_2)$, $\Delta P=p_1+p_2$, 
where $p_1$, $p_2$ are the momenta of two particles. 
Next, consider the case $k=0$. Clearly in this case 
the number of vacancies at the real axis is $L/2$ 
and the low-energy excitations correspond to the 
one particle and one hole. The energy and the momentum 
of this state are exactly the same as in the previous 
case. Now $p_1$ and $p_2$ are the momenta of the particle 
and the hole. Finally at $k=-1$ we have the state with
two holes which has the same energy and the momentum 
as for the two particles.

Finally, let us note that the particle states plays 
an essential role in the derivation of the so called 
Luttinger liquid relation \cite{H} for the XXZ- spin chain. 
In fact one should take into account that the extra numbers 
of particles at the left and the right Fermi- points 
can be positive or negative. Clearly in the case of the 
positive sign one needs the complex roots (particles) 
to derive the Luttinger liquid relation completely.

\vspace{0.6in}

{\bf 5. Conclusion}

\vspace{0.2in}

In the present paper we stressed that for the XXZ- spin chain 
there is a large amount of the eigenstates connected with the 
complex roots with the imaginary part equal to $\pi/2$. 
We have studied the evolution of the states with the complex roots
with varying the anisotropy parameter. We have determined 
the XXX- spin chain limit of these states. 
We also proposed the particle-hole symmetry relations 
between the quantum numbers of the state corresponding to the 
two different pseudovacuum states. 
Our results are important both for the determination of the 
low-lying particle-hole excitations and for the derivation 
of the Luttinger liquid theory relation for the XXZ- spin chain.


\begin{thebibliography}{99}

\vspace{0.2in}

\bibitem{Bethe}
H.Bethe, Z.Physik, 71 (1931) 205. 

\bibitem{O}
L.Hulthen, Arkiv.Mat.Astron.Fysik. 26A(1938) No 11;~ 
R.Orbach, Phys.Rev. 112 (1958) 309;~ 
C.N.Yang, C.P.Yang, Phys.Rev. 150 (1966) 321. 

\bibitem{Takahashi}
M.Takahashi, Progress Theor.Phys. 46 (1971) 401. 

\bibitem{V}
C.Destri, H.J.de Vega, Nucl.Phys.B 504 (1997) 621. 

\bibitem{FT}
L.Faddeev, L.Takhtajan, Phys.Lett.A 85 (1981) 375. 

\bibitem{G}
M.Gaudin, ``La fonction d'onde de Bethe'', Masson, 1983. 

\bibitem{H}
F.D.M.Haldane, Phys.Rev.Lett. 47 (1981) 1840;~ J.Phys.C 14 (1981) 2585~;
Phys.Rev.Lett. 45 (1980) 1358.  

\end{thebibliography}
\end{document}